\documentclass{article}
\usepackage{graphicx}
  
\begin{document}

\title {Contribution of a nearby Pulsar to Cosmic Rays observed at Earth}
\author{A. Bhadra\thanks{Email address: aru\_bhadra@yahoo.com}  \\ 
University of North Bengal, Siliguri 734430 
INDIA \\
}
\date{}
\maketitle 
\begin{abstract}
Contribution of nearby pulsars to the cosmic rays observed at Earth has been studied. It is found that the experimental bound on amplitude of cosmic ray anisotropy may produce significant constraint on the efficiency of converting pulsar rotational energy to emitted particles kinetic energy. Cosmic ray fluxes from two well known nearby gamma ray pulsars, namely the Vela and Geminga pulsars, are estimated. The analysis suggests that observed bound on cosmic ray anisotropy restricts the contributions of the Vela and the Geminga pulsars to at most $1$ \% of the observed cosmic rays below the knee. 
\end{abstract}


\section{Introduction}
Pulsars are widely considered to be natural sites for acceleration of charged particles (e.g. [1-12]). Along with the Supernova explosions and Gamma Ray bursts, they are regarded as the most probable sources of ultra high-energy cosmic rays within the galaxy. From the point of view of available energy, pulsars are even a more promising source because the rotational energy at birth can be more than ten times of the energy released in a typical supernova explosion. It also has the necessary conditions for accelerating particles to ultra high energies. Moreover observation of pulsed and/or steady flux of electromagnetic radiations from radio to gamma ray energies provides direct evidence that some pulsars are site of energetic particles of at least several TeV. Another important feature is the maximum energy ($E_{max}$) attainable by a particle in the acceleration process. In the case of SNR this is at most $10^{15}$ eV [13]. On the other hand according to the Hillas condition [14], the maximum energy of a particle of charge Ze that can be contained near the light cylinder of a pulsar of angular speed $\Omega \; rad \;s^{-1}$, radius $R$ and with the surface magnetic field $B_{s}=B_{12} \times 10^{12}$ Gauss is
\begin{equation}
E_{max}=3.4 \times 10^{17} Z B_{12} \left( \frac{\Omega}{10^3 \; s^{-1}} \right)^2 \left( \frac{R}{10^6 \; cm} \right)^3 \; eV
\end{equation}   
which shows that the maximum possible energy can be very large, for fast rotating pulsars $E_{max}$ even could reach around $100$ EeV [1,7], the highest energy cosmic ray particles observed so far. A point that usually has been raised against the pulsar model of origin of cosmic rays is that the derived spectrum of cosmic rays from pulsars is much flatter than the observed spectrum. However, such derivation is based on the assumption that the production spectra from all pulsars have the same slope and $E_{max}$. It has been shown recently [9,10] that the expected cosmic ray spectrum in the energy range from PeV to EeV coincides with the observation if the distribution of pulsar initial periods is similar to the Gamma distribution [9] or if the log of initial pulsar periods and surface magnetic fields are given by the Gaussian distribution [10]. \\
An important characteristics of cosmic rays is that they are highly isotropic; the amplitude of anisotropy is less then $10^{-2}$ below the knee energy region [15]. While estimating contribution of a pulsar such isotropic nature of cosmic rays has to be respected. But this point has not been taken into consideration in several investigations. In most of the recent efforts of estimating cosmic rays from pulsars collective contribution of all galactic pulsars are studied [7-10]. Hence it is not clear from those analysis whether the adopted cosmic ray production scenario in pulsars is consistent with the observed isotropy of cosmic rays or not. Here it is worthwhile to mention that the observed upper limit of cosmic ray anisotropy leads to some important conclusions on the supernova remnant (SNR) model of origin of cosmic rays as well as on their propagation. For instance in a recent work [16] it is shown that the observed cosmic ray isotropy does not support the idea the that only a small fraction of supernova remnants provide the main contribution to the cosmic ray flux in the knee region [17]. In another work [18] it is found that the SNR origin model of cosmic rays is compatible with the observed upper limit of cosmic ray anisotropy if the cosmic ray propagation is by diffusion with re-acceleration; the plain diffusion model of propagation gives too large anisotropy. Thus the study of individual contribution of pulsars, particularly from nearby pulsars, is of great significance. In the context of single source model of the knee [19] contribution of nearby pulsars to the cosmic ray spectrum at knee region has been studied recently [20] where parameters are chosen suitably so that the contribution of the pulsars becomes significant at the knee energy of the spectrum. But isotropic nature of cosmic rays was not incorporated in those studies. In the present work cosmic ray flux from nearby pulsars are estimated respecting the observed isotropy of cosmic rays. \\
The plan of the work is as follows. After giving a short review of the acceleration of charged particles by pulsars, we obtain the cosmic ray spectrum of charged particles produced by a pulsar in sec. II. The contribution of a nearby pulsar to the cosmic rays at the Earth is estimated in section III. In section IV constraint on the contribution of cosmic rays from a point source that may arise from the observed isotropy of cosmic rays are discussed. Fluxes of cosmic rays from two nearby gamma ray pulsars, the Geminga and Vela pulsars, are estimated in section V. Finally the results are concluded in section VI. 
  
\section{Acceleration of nuclei by pulsars}
Pulsars are generally believed to be rotating neutron stars [21].  The rotation of the magnetic dipole of pulsars generates huge electrical differences between different parts of its surface and as a result the rotating neutron star should be surrounded with charged plasma, which is called a magnetosphere. Since the moment of inertia of a neutron star is around $10^{45}\; erg \; s^2$, a millisecond pulsar has a rotational energy $E=\frac{1}{2} I \Omega^2 \sim 10^{52}$ ergs. A fraction of such a huge rotational energy of a pulsar may be converted to the kinetic energy of the particles those present in the magnetosphere. \\
The pulsar magnetosphere is usually considered to be composed of electron-positron pairs. They are expected to be produced to get rid of electric fields that are parallel to magnetic fields at a region where net charge density is different from the Goldreich-Julian density [1]. However, hadronic component also may exist [1-12] in magnetosphere. Ions in the magnetosphere of a neutron star are most likely to come from the surface of the star. The surface of the neutron star is supposed to compose of heavy nuclei, most probably of iron nuclei [12]. The binding energy of these surface nuclei is, however, not accurately known. In the case of small binding energy ($<1 \; keV$), thermionic emission of ions from the surface of the star is possible even for the old stars which may give practically free outflow of ions in the magnetosphere. But if the binding energy of surface atoms is large ($2-3 \; keV$), thermionic emission can take place only for young pulsars (surface temperature $>10^{6o} C$). The local temperature of the polar cap can be higher due to the bombardment by particles of the magnetosphere and thus thermionic emission still may occur in pulsars of intermediate age. Surface atoms also can be stripped off by the strong electric field of the polar cap of a pulsar [4].  \\   
These nuclei can be accelerated by pulsars through large potential drop associated with strong electric field parallel to the pulsar magnetic field. Several detailed mechanisms  have so far been suggested for accelerating particles by pulsars including the popular polar gap [4], and the outer gap [5] models. In the former model, acceleration of particles takes place in the open field line region above the magnetic pole of the neutron star whereas in the case of outer gap model it occurs in the vacuum gaps between the neutral line and the last open line in the magnetosphere. Thus, the region of acceleration in the polar gap model is close to the pulsar surface, while the same in the outer gap model is close to the light cylinder. It has also been suggested that pulsars may accelerate protons and heavy nuclei by converting their rotational energy to particle kinetic energy via a relativistic MHD winds near the light cylinder [3,7]. In this process particles may gain energy from the largest potential drop associated with a neutron star. In the present investigation, however, we restrict our discussion within somewhat general framework of electromagnetic acceleration process without focusing on any specific model. 

\subsection{Maximum energy}
A large-scale electric field in the magnetosphere is expected to develop due to the rotating magnetic dipole of the pulsar. Neglecting the influence of the pulsar magnetosphere, the maximum electrostatic potential at time $t$ is 
\begin{equation}
\phi_{m}= \frac{1}{2c^2} B_{s} \Omega^2(t) R^3
\end{equation}
But the whole potential drop may not be available for the particle acceleration due to development of electron-positron pair cascades (originate from curvature radiation photons emitted by a seed electron accelerated by the drop) in opposite direction which quickly shuts off the potential drop [5, 12, 22]. \\
If $f$ fraction of the $\phi_{m}$ is responsible for the acceleration process, the energy of the accelerated nuclei would be
\begin{equation}
E^{cr}= 1.7 \times 10^8 f Z B_{12} \left(\frac{\Omega(t)}{10^3 \; s^{-1}} \right)^2 \left( \frac{R}{10^{6} \; cm} \right)^3 \; GeV.
\end{equation}
In general $f$ may depend on pulsar period and magnetic field. On the other hand if we consider the acceleration scenario in which magnetic field energy near the light cylinder is transferred to particle kinetic energy [3,7], the magnetic energy per ion at the light cylinder is $E= \frac{B_{lc}^{2}}{8 \pi n_{GJ}}$ where $B_{lc}=B_{s}R^{3}\Omega^{3}/c^{3}$ is the magnetic field at the light cylinder and $n_{GJ}(r)=B_{s}R^{3} \Omega /(4 \pi Ze c r^{3})$ is the Goldreich-Julian density. Thus the numerical value of the magnetic energy per ion becomes the same as Eq.(3) with $f=1$. 

\subsection{Energy spectrum}
In the acceleration picture mentioned above it is natural to consider that at any evolutionary stage of the pulsar, all the particles are accelerated under the same potential difference and hence should attain the same maximum energy $E^{cr}$. With the spin down of the pulsar, attainable energy of the accelerated particles changes. If we assume that the $\xi$ fraction of total rotational energy loss of a pulsar goes to accelerate nuclei then one can write the rate of number of accelerated particles at any moment
\begin{equation}
\frac{dN}{dt}=\xi \frac{\dot{E^{rot}}}{E^{cr}} \; \; s^{-1}
\end{equation}
However, the charged density follows from the above equation may not be always less than the Goldreich-Julian density ($\rho \approx -\Omega B/(2\pi c)$) [8]. But the expression for the GJ density is valid only in the co-rotation portion of the magnetosphere [1,8] whereas the most likely site of production of high-energy cosmic rays is near the light cylinder. If $\eta$ fraction of total rotational energy loss is due to the emission of magnetic dipole radiation, then  
\begin{equation}
\eta I \Omega \dot{\Omega}=\frac{B_{s}^2 R^6 \Omega^4}{6 c^3}
\end{equation}
So one can express the rate of change of rotational energy of a pulsar at any instant t as 
\begin{equation}
\dot{E^{rot}} = 3.9 \times 10^{42} \eta^{-1} B^{2}_{12} \left(\frac{\Omega (t)}{10^3 \; s^{-1}}\right)^4 \left(\frac{R}{10^6 \; cm}\right)^6 \; GeV \; s^{-1}
\end{equation}
The evolution of the pulsar period is given by the expression
\begin{equation}
P^2(t_2)  = P^2(t_{1}) +4.8 \times 10^{-16} (t_2-t_{1}) B^{2}_{12} \left(\frac{R}{10^6 \; cm}\right)^6 \left(\frac{\eta I}{10^{45} \;g \; cm^2}\right)^{-1} 
\end{equation}
Here $P(t_2)$ and $P(t_{1})$ are the period of pulsar in seconds at time $t_2$ and $t_{1}$ respectively. The rate of change of pulsar period at a particular instant t is followed from the above equation which is given by
\begin{equation}
\dot{P}(t)  = 2.4 \times 10^{-16} \frac{B^{2}_{12}}{P(t)} \left(\frac{R}{10^6 \; cm}\right)^6 \left(\frac{\eta I}{10^{45} \;g \; cm^2}\right)^{-1} 
\end{equation}
With the evolution of the pulsar period, the energy of the accelerated particles changes according to Eq.(3) through Eq. (7). The resulting energy spectrum of the accelerated particles thus reads as 
\begin{equation} 
\frac{dN}{dE}  = \xi \frac{1.9 \times 10^{45}}{E}  \left(f Z B_{12} \left(\frac{R}{10^6 \; cm}\right)^3 \right)^{-1} \left(\frac{ \eta I}{10^{45} \;g \; cm^2}\right)  \; GeV^{-1}
\end{equation}
Apparently the energy spectrum of the produced accelerated particles has the $E^{-1} $ behavior. But since the production times for particles of different energies are different, the observed spectrum may not go simply as $E^{-1}$ [8].   

\subsection{Escape from the nebula}
A young pulsar is generally encircled by the remnant of the pre supernova star. In fact most gamma ray pulsars have plerions as well as they have associated supernova remnants. Only the oldest pulsars such as Geminga or PSR1055-52 show no sign of having plerions. So pulsar accelerated nuclei will inject into the nebula. Initially the accelerated nuclei will interact with the dense matter of the expanding envelope. The interactions, however, become negligible when the column density is very much less than the interaction length of the energetic nuclei in the medium. The column density of the matter in envelope offered to the injected high energy nuclei after $t$ years from the birth of the pulsar is approximately
\begin{equation}
\Sigma(t) \simeq \frac{M_{neb}}{4 \pi R^{2}_{neb}}= 50 \left(\frac{t}{1 \; year}\right)^{-2} \left( \frac{M_{neb}}{10 \; M_{\odot}} \right)^{2}\left( \frac{E_{SN}}{10^{51} \; ergs}\right)^{-1} \; g \; cm^{-2} 
\end{equation}
where $M_{neb}$ and $R_{neb}$ stand for mass and effective radius during the free expansion phase of the envelope respectively. The radius of the envelope varies as $R_{neb} \simeq R_{o} + v \; t$ where $R_{o}$ is the radius of pre-supernova star ($R_{o} \preceq 10^{14}$ cm) and the velocity of the free expansion $v \sim (2 E_{SN}/M_{neb})^{1/2}$, the observations suggest that the free expansion velocity of Crab nebula is around 2000 $KM \; s^{-1}$. The condition for iron nuclei to pass through the envelope without major losses is that $\Sigma \preceq 100 \; g cm^{-2}$. Thus the energetic nuclei can escape the remnant without significant losses shortly after the explosion. \\
However, energetic nuclei injected into the envelope may be trapped by the magnetic field of the nebula. Outside the light cylinder the azimuthal component of the magnetic field dominates over the radial field and is given by 
\begin{equation}
B(r) \simeq \sqrt{\sigma} B_{s}\left( \frac{R}{r_{lc}} \right)^{3} \frac{r_{lc}}{r}
\end{equation}
where $\sigma$ is the ratio of the magnetic field flux to the particle energy flux at shock radius r. For the Crab pulsar $\sigma$ is estimated to be $0.003$ [23] whereas for the Vela pulsar the value of $\sigma$ is close to $1$ [24]. According to the standard picture of development of pulsar envelope [25] at the initial stage envelope expands freely with velocity $v$. At this stage the radius of the envelope increases with time roughly as $6.3 \times 10^{15} t$ cm where $t$ is in years. With time the envelope slowly begins to decelerate by transferring its energy to the ambient medium. When the envelope reaches the Sedov phase, which occurs when the matter density of the nebula becomes comparable to that of the surrounding medium and begins at $t_{sedov} \simeq R_{sedov}/v $ where the radius of the envelope at Sedov phase $R_{sedov} \simeq \frac{1}{4 \pi} \left( \frac{M_{env}}{n}\right)^{1/2}$ with $n \simeq 0.3 \; g cm^{-3}$ being the density of the surrounding medium (typically $t_{sedov} \sim 2000$ years), the velocity of envelope can be approximated as $v_{sedov}=v (t_{sedov}/t)^{3/5}$ [26]. Consequently the radius of the envelope goes with time as  $r_{neb} \simeq v t_{sedov}^{3/5} t^{2/5}$. During this stage the strength of the magnetic field in the outer part of the envelope takes a value around $3$ to $5 \mu G$, the typical value for the interstellar medium. Accordingly, the Larmor radius ($r_{L}=E/(ZeB)$) in the envelope becomes about $10^{18} \left( \frac{E}{1 \; PeV} \right) \frac{1}{Z}$ cm. Thus, nuclei of energy even $1$ PeV should be trapped by the magnetic field of the nebula. However, it is known that several instabilities develop in the process of confinement of charged particles by magnetic fields and the high- energy nuclei will finally escape along the field lines of the irregular fields [8]. \\
The energetic nuclei will propagate diffusively in the envelope before escaping into the interstellar medium provided diffusion time $ \tau_{diff}$ ($\equiv r^{2}/4D(E) $, $D(E)$ is the diffusion coefficient) for traversing the radial distance $r_{neb}$ is much larger than $\tau=r_{neb}/c$, the time required for straight line propagation. When the energy of particles is too high they move on almost rectilinear trajectories. The pulsar accelerated particles will escape from the nebula when the mean radial distance traveled by the particles becomes comparable with the radius of the nebula at the time of escaping. The mean radial distance traveled in diffusive propagation in time $t$ is [27] 
\begin{equation}
< r_{rad}> = 2 \sqrt{\frac{ D(E) t}{\pi  }}  
\end{equation}
where $D(E)$ is assumed as time independent. Such an assumption holds, at least as first approximation, only after the nebula enters the Sedov phase, as thereafter the magnetic field of the nebula does not change appreciably with time. The Diffusion coefficient for relativistic particles is defined by the relation $D(E) = \frac{1}{3} c \lambda_{d}$, where $\lambda_{d}$ is the mean scattering length by magnetic irregularities. The problem is that $\lambda_{d}$ is not known with good accuracy. Theoretical limit on $\lambda_{d}$ is that it cannot be smaller than the Larmor radius [28]. Thus the minimum value of diffusion coefficient (so-called Bohm diffusion constant) corresponding to maximally turbulent medium is given by 
\begin{equation}
D=\frac{1}{3} c r_{L} \;.
\end{equation}
But after the nebula enters into the Sedov phase the magnetic field fluctuations is more likely to represent by the Kolmogorov spectrum [12]. In that case the diffusion coefficient is well described by the relation 
\begin{equation}
D(E) =\frac{1}{3} r_{L}^{1/3} l_{cell}^{2/3} \; .
\end{equation}
where $l_{cell}$ represents the characteristic size of the magnetic cells (irregularities) in the nebula which is obviously less than the radius of the nebula but can be within an order of that, the characteristic size of filamentary structure of the Vela supernova remnant as revealed from multiwave length studies indicates such possibility.  \\
It is evident from the foregoing discussion that mean storage time of cosmic rays of different energies in the nebula depends on the choice of the diffusion coefficient. But even for the Bohm diffusion, which gives maximum confinement, mean storage time of a PeV proton in the nebula is only few hundred years whereas the same for a PeV iron nucleus is few (typically $6$ to $10$) thousand years.  \\

\section{Cosmic ray flux from a pulsar to be observed at Earth}
The diffusion process governs the propagation of accelerated charged nuclei from the source. Usually the diffusion scenario is considered as Gaussian. 
Neglecting the effect of energy gained/losses during propagation, convection, losses of nuclei by collision and nuclear interactions, the diffusion equation is given by
\begin{equation}
\frac{d N}{d t}= \Delta . (D(E) \Delta N(r,t,E) + Q(r,t,E) 
\end{equation}   
where $Q(r,t,E)$ represents the source term. The Green's function for the above equation is given by
\begin{equation}
G = \frac{1}{8(\pi D t)^{3/2}} exp[-r^2/(4D t) ] \; cm^{-3}
\end{equation}
Here $t$ is the time when cosmic ray of energy $E$ escapes from the pulsar nebula. Thus the intensity of cosmic rays of energy $E$ at a distance $r$ from a pulsar would be
\begin{eqnarray}
I_{cr}(r,E)= \xi 2.6 \times 10^{52}\left(f Z B_{12} \Omega (t) \left(\frac{R}{10^6 \; cm}\right)^3 \right)^{-2} \left(\frac{\eta I}{10^{45} \;g \; cm^2}\right) G(r, \tau)  \nonumber \\
cm^{-2}s^{-1}sr^{-1} GeV^{-1}
\end{eqnarray}  
where c is the speed of light. 

\section{Anisotropy}
The anisotropy predicted for galactic cosmic ray sources relies on the model for the production of cosmic rays and for their propagation. The production of cosmic rays depends on the nature of the sources whereas the diffusion process due to scattering from minor irregularities in the field governs the propagation of accelerated charged nuclei from the source to Earth. The resultant amplitude of anisotropy is given by [29]
\begin{equation}
\delta=\frac{\lambda_{d}}{I(E)} \big{\vert} \frac{\partial I (E)}{ \partial r}  \big{\vert}
\end{equation}
where I is local cosmic ray intensity and $\partial I/ \partial r$ is the intensity gradient. Employing the simplified model of diffusion as mentioned in previous section, one gets  
\begin{equation} 
\delta = h(E) \frac{3r}{2c t (E)}  
\end{equation}
Here $h(E)$ denotes the ratio of the cosmic rays of energy $E$ from the source to the total observed flux of cosmic rays at the same energy from all sources. The energy dependence of $\delta$ relies mainly on $h(E)$. But emission time and hence $t$ also can be different for cosmic rays of different energies depending on the nature of source. A nice feature of the expression (19) is that it does not depend on the diffusion coefficient. However, if a different propagation scenario, such as the so-called anomalous diffusion [30], is adopted then the expression for anisotropy amplitude might depend on the diffusion coefficient. Here it should be mentioned that the numerical value of diffusion coefficient at concerned energy range is quite uncertain. Another important point is that once the contribution of the source to the total cosmic ray flux is fixed, there is no adjustable parameter left in the expression (15). \\
As the observations show that cosmic rays are highly isotropic, $\delta$ is restricted to a small value. For example harmonic analysis of the right ascension distribution of cosmic rays gives amplitude of anisotropy upto the knee region is around $10^{-2}$ to $10^{-3}$ [15]. The measurements at the higher energy region suffer from the low flux of cosmic rays and thus $\delta$ is less restricted there. While estimating flux from a point source constraint imposes by the observed anisotropy through the Eq.(19) has to be respected which in turn may restrict the parameter $\xi$. The phase of the amplitude of first harmonic is also a quantity of significance; the contribution of a point source to the total cosmic ray intensity needs to be consistent with the observed phase of the amplitude [15].  

\section{Application: Cosmic rays from two nearby gamma ray pulsars}
In a recent work [8] it has been shown that Gamma ray pulsars of the Galaxy could contribute a significant fraction of the total cosmic ray flux at and above knee energy region. However, the contribution of the nearby sources has not been considered there. Effect of observed bound of cosmic ray anisotropy on the  parameters involved in acceleration of cosmic rays by pulsars, if any, is expected to reveal from the study of contribution of cosmic rays from the nearby pulsars. For pulsars of large distances the effect of different emission (form pulsars) times of cosmic ray particles having different energies on the observed cosmic ray spectrum are not significant. This is because in such cases the propagation time is large compared to the life time of the pulsars. The effect of different time of emission could be important only for nearby pulsars. In the following cosmic ray fluxes from the two nearby gamma ray pulsars, namely the Vela and the Geminga, are considered. The present analysis is restricted to gamma ray pulsars only as the emission of gamma rays is a definite signature of presence of high-energy particles. We take  $D(E)=D_{o} (E/Z)^{\delta}\; cm^2 s^{-1}$ (where $E$ is in GeV) with $D_{o}=1.4 \times 10^{28}$ for galactic magnetic field and $\delta=1/3$ [31] . 

\subsection{The Vela Pulsar}
Vela was found by SAS 2 as the brightest object in the gamma ray sky. It is comparatively close to the Earth and so the surrounding nebula is well studied. Its characteristic age is around $10,000$ years with periodicity $89.3 $ ms, slow down rate is $1.25 \times 10^{-13} $, and surface magnetic field is around $3.4 \times 10^{12}$ Gauss [32]. The distance of the object from the Earth is around $500$ pc [33] though recent works suggest a smaller value of $300$ pc [34].  

\begin{figure}[ht]
\centering 
\includegraphics[width=0.75\textwidth,clip]{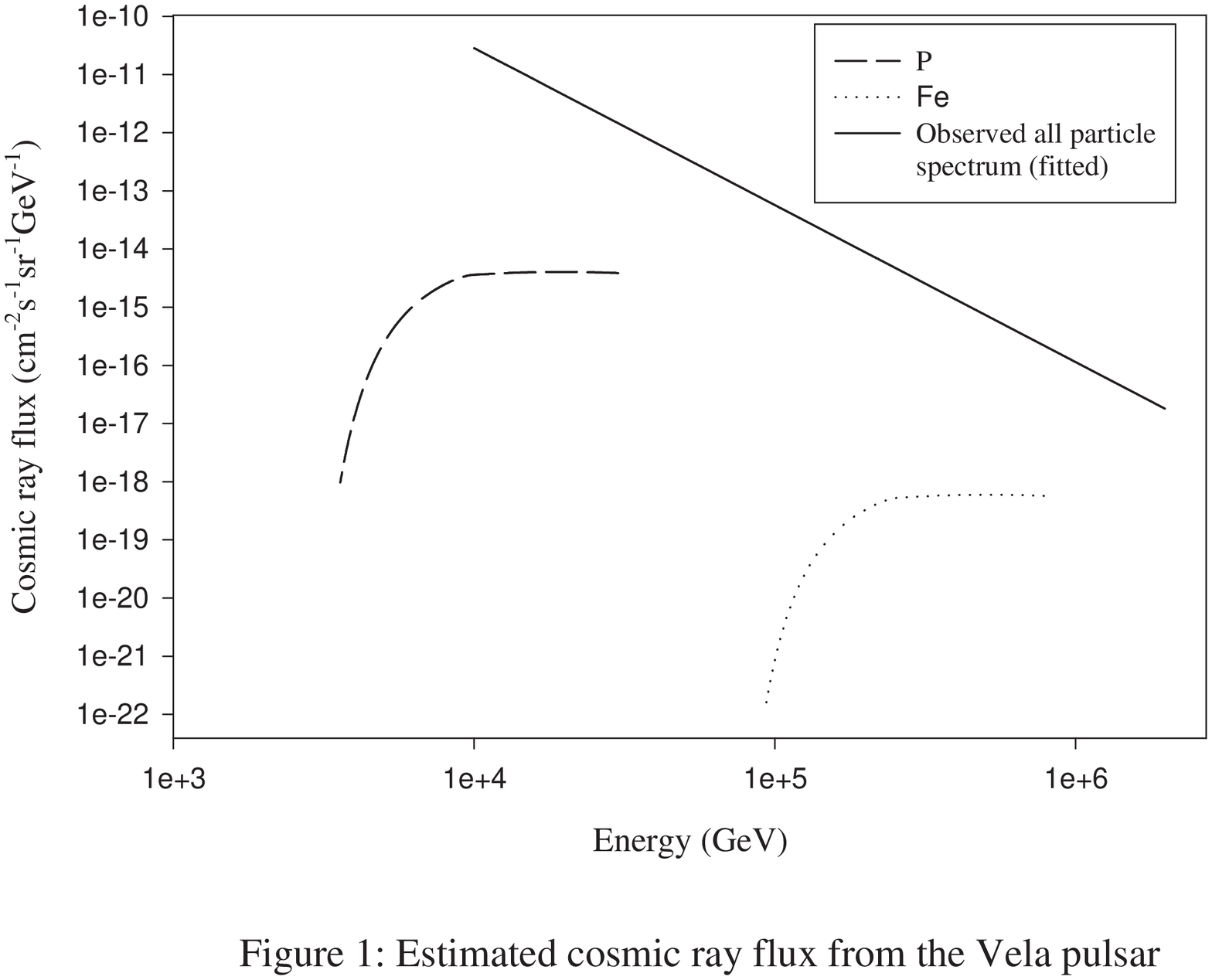}
\caption[Estimated Cosmic Ray flux from the Vela pulsar]{Estimated Cosmic Ray flux from the Vela pulsar}
\end{figure}

The observed pulsar period and slow down rate are consistent with both the Eqs. (5), and (8) for $ \left(\frac{R}{10^6 \; cm}\right)^6  \left(\frac{\eta I}{10^{45} \;g \; cm^2}\right)^{-1} = 0.24$. The initial pulsar period is estimated as $26.3$ ms from Eq.(7). In the framework of the self-sustained outer gap model [35] in which $f$ is given by
\begin{equation}
f=1.07 \times 10^{-4} \left(\frac{\Omega(t)}{10^3 \; s^{-1}} \right)^{-52/21} B_{12}^{-8/7} \;,
\end{equation}
the maximum energy would be $ 30 \; Z$ TeV. Since the pulsar is relatively young in the case of Bohm diffusion it is likely that the pulsar accelerated cosmic rays still remain confined in the nebula. But if the characteristic size of the magnetic irregularities in the nebula is larger than the Larmor radius then the diffusion coefficient will be given by the Eq.(14) and assuming $l_{cell}$ is of the same order to the radius of the nebula  (corresponds to minimum confinement) it is found that the mean storage time of a $30$ TeV proton is around $3000$ years; the same for a $100$ TeV iron nuclei is about $6000$ years. Resulting spectra of cosmic rays from the pulsar to be observed at earth are shown in figure 1 for both proton and iron primaries assuming $\xi=0.01$ and $0.001$ respectively. The parameter $\xi$ is so chosen to respect the observed  isotropy of cosmic rays. The observed all particle cosmic ray (best fitted) spectrum is also given in the figure for comparison.

\subsection{The Geminga pulsar}
Another pulsar in the area is Geminga (PSR J0633+1746), which is discovered by the SAS-2 group [36] and later confirmed by the COS-B group [37], about $150$ pc away from the Earth. Its radial velocity is unknown, but if it were 200 KM/sec, it could have been within $100 \; pc$ of Earth at 340,000 years ago. Geminga is a unique object: a highly compressed, spinning neutron star which does not emit radio waves like the other well-known pulsars. Yet it is a powerful source of pulsating gamma-rays and X-rays. Geminga is now known to be a rotation-powered pulsar with period $P = 0.237$ s, $\dot{P}=1.0975 \times 10^{-14}$ and surface magnetic field $B= 1.6 \times 10^{12}$ G [38]. 

\begin{figure}[ht] 
\centering
\includegraphics[width=0.75\textwidth,clip]{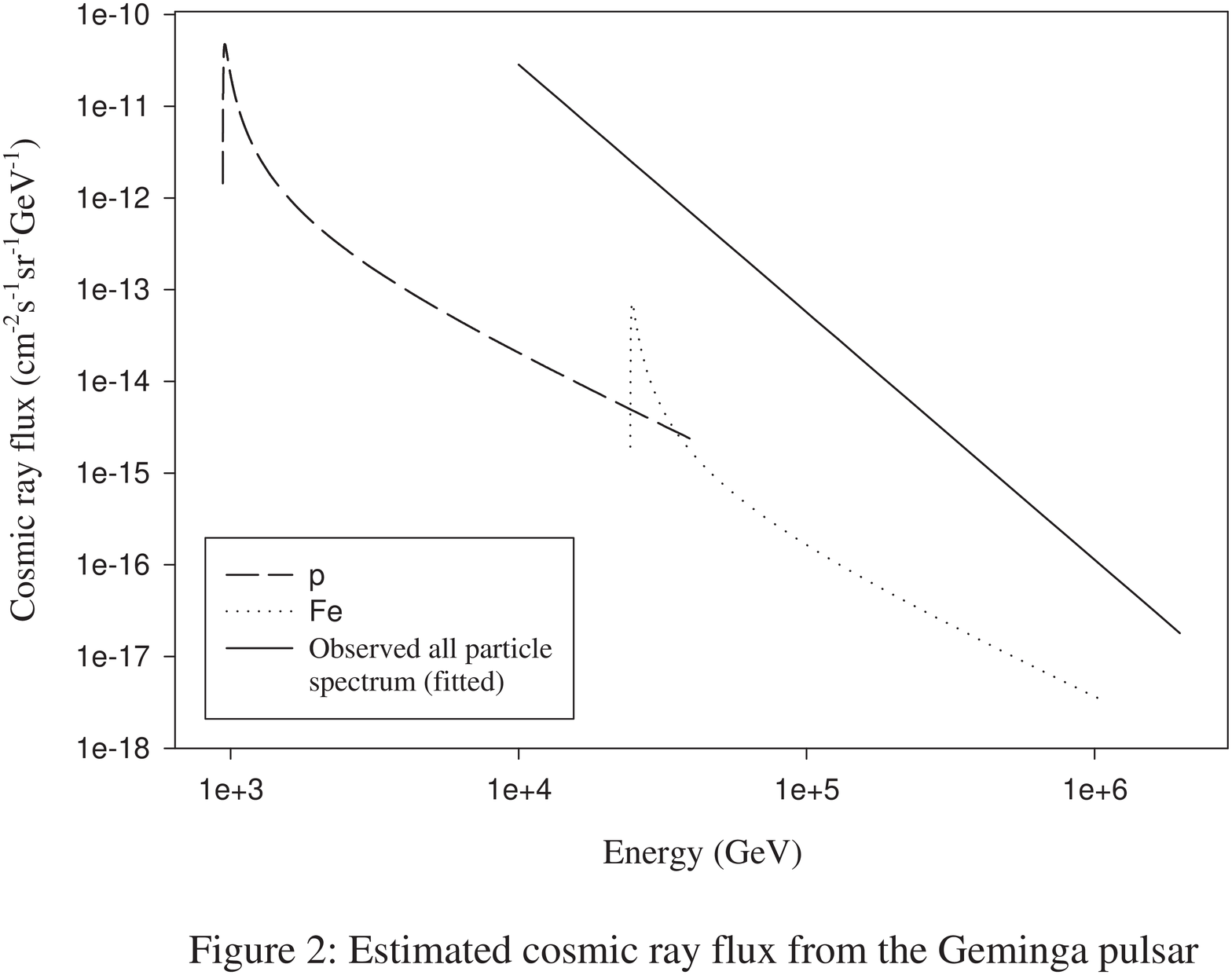}
\caption[Estimated Cosmic Ray flux from the Geminga pulsar]{Estimated Cosmic Ray flux from the Geminga pulsar}
\end{figure}

Since Geminga is a relatively old pulsar, presence of ions in its magnetosphere is certainly questionable. However, observation of high-energy gamma rays from this pulsar indicates the presence of energetic particles in its atmosphere. Besides the non-observation of radio waves from the pulsar suggests that these high-energy particles may not be electrons. Hence it is likely that ions are present in its magnetosphere; at least one cannot rule out this possibility. Even if sufficient ions are not available in the pulsar atmosphere presently, one needs to consider the high energy particles which were produced at earlier stages. According to Eq.(7), the initial pulsar period is 36.6 ms. Here also we take $ \left(\frac{R}{10^6 \; cm}\right)^6 \left(\frac{\eta I}{10^{45} \;g \; cm^2}\right)^{-1} =0.24$ which follow from Eq. (5) and is consistent with Eq. (8). The maximum energy of the energetic nuclei thus would be only $4 Z \times 10^{13}$ eV when $f$ is given by Eq.(20). Taking $\xi=0.1$ so that observed isotropy is mostly obeyed, we estimate the spectra of cosmic rays from the pulsar to be observed at earth for proton and iron primaries which are shown in figure 2. Observed (best fitted) all particle spectrum [39] is also presented for comparison. It is worthwhile to mention that for Geminga the nature of diffusion process of cosmic rays in the supernova envelope has little effect on the resultant (observed) spectra because of its age. 

\section{Discussion}
In the present work contribution of a nearby pulsar to cosmic rays observed at earth has been studied respecting the observed isotropy of cosmic rays. It is found that the parameters involved in the cosmic ray generation process in pulsars, particularly  the efficiency of converting pulsar rotational energy to emitted particles kinetic energy may be severely constrained by the observed limit on cosmic ray anisotropy. Consequently production of high-energy gamma rays through collision of cosmic rays with the ambient matter would be also restricted [40]. \\
Cosmic ray fluxes from two well-known gamma ray pulsars, namely the Vela and Geminga pulsars, are obtained. It is found that owing to the observed bound on cosmic ray anisotropy the contributions of the Vela and the Geminga pulsars are limited to at most $1$ \% of the total cosmic ray intensity below the knee. This  means the Vela and the Geminga pulsars do not contribute substantially to the total cosmic ray intensity. Such a conclusion is consistent with the observed phase of the amplitude of first harmonic; below the knee energy maximum flux of cosmic rays (phase of the anisotropy) is observed from the second quadrant of the Galaxy [15], while both Vela and Geminga are in the third quadrant. \\
Another interesting observation is that the cosmic ray spectra to be observed at Earth from both the pulsars, the Geminga and the Vela, are found noticeably different than the production spectra which is mainly due to combined effect of different generation (and emission) time of pulsar accelerated cosmic rays of different energies and their propagation. 

\section*{ Acknowledgment}
The author is greatful to an anonymous referee for helpful suggestions.

\end{document}